\documentclass[aps,pre,epsf,superscriptaddress,twocolumn,showpacs]{revtex4}
\usepackage{titlesec}
\usepackage{graphicx}
\usepackage{epsfig}
\usepackage{dcolumn}
\usepackage{bm}
\usepackage{braket}
\usepackage{color}
\usepackage{amsmath}
\begin{document}

\title{Resonant quantum dynamics of few ultracold bosons in periodically \\driven finite lattices}

\author{S.I. Mistakidis}
\affiliation{Zentrum f\"{u}r Optische Quantentechnologien,
Universit\"{a}t Hamburg, Luruper Chaussee 149, 22761 Hamburg,
Germany}
\author{T. Wulf}
\affiliation{Zentrum f\"{u}r Optische Quantentechnologien,
Universit\"{a}t Hamburg, Luruper Chaussee 149, 22761 Hamburg,
Germany}
\author{A. Negretti}
\affiliation{Zentrum f\"{u}r Optische Quantentechnologien,
Universit\"{a}t Hamburg, Luruper Chaussee 149, 22761 Hamburg,
Germany}\affiliation{The Hamburg Centre for Ultrafast Imaging,
Universit\"{a}t Hamburg, Luruper Chaussee 149, 22761 Hamburg,
Germany}
\author{P. Schmelcher}
\affiliation{Zentrum f\"{u}r Optische Quantentechnologien,
Universit\"{a}t Hamburg, Luruper Chaussee 149, 22761 Hamburg,
Germany} \affiliation{The Hamburg Centre for Ultrafast Imaging,
Universit\"{a}t Hamburg, Luruper Chaussee 149, 22761 Hamburg,
Germany}

\date{\today}

\begin{abstract}
The out-of-equilibrium dynamics of finite ultracold
bosonic ensembles in periodically driven one-dimensional optical lattices is
investigated. Our study reveals that the driving enforces the bosons
in different wells to oscillate in-phase and to exhibit a
dipole-like mode. A wide range from weak-to-strong driving
frequencies is covered and a resonance-like behaviour of the
intra-well dynamics is discussed. In the
proximity of the resonance a rich intraband excitation spectrum is
observed. The single particle excitation mechanisms are studied
in the framework of Floquet theory elucidating the role of the
driving frequency. The impact of the interatomic repulsive
interactions is examined in detail yielding a strong influence on
the tunneling period and the excitation probabilities. Finally, the
dependence of the resonance upon a variation of the tunable
parameters of the optical lattice is examined. Our analysis is
based on the ab-initio Multi-Configuration
Time-Dependent Hartree Method for bosons.\\

\end{abstract}

\maketitle

\section{Introduction}

Ultracold atomic quantum gases in optical lattices have reached an
unprecedented degree of control providing
direct experimental access to a plethora of non-equilibrium
phenomena \cite{Goldman,Goldman1,Morsch1,Bloch}. This
control includes the modulation of the interparticle
interactions via confinement-induced, magnetic and optical Feshbach resonances
\cite{Olshanii,Grimm,Santos,Inouye,Kohler,Chin}, the design of
arbitrarily shaped optical traps with variable lattice depths, 
and the ability to move time-periodically or even accelerate the
entire lattice structure. This level of control and accuracy over
the system parameters has opened the possibility to simulate and
study quantum many-body phenomena in part inspired from
condensed matter physics \cite{Lewenstein1}. For instance, when
accelerating an optical lattice, representative processes are Bloch-oscillations
\cite{Choi,Dahan,Morsch,Peik,Cristiani}, Wannier-Stark ladders
\cite{Wilkinson,Niu}, Landau-Zener tunneling \cite{Niu,Cristiani}
and photon assisted tunneling \cite{Sias}, to name only a few. A
promising technique is the lattice shaking which has been used in
order to address e.g., the coherent control of
the superfluid to Mott insulator phase transition \cite{Eckardt},
parametric amplification of matter waves \cite{Gemelke}, four-wave
mixing \cite{Hilligsoe,Lopes}, topological states of matter
\cite{Zheng}, hybridized band structure \cite{Gemelke,Lignier}, and
even the engineering of artificial gauge fields \cite{Struck}. More
recently it has been shown \cite{Parker,Choudhury} that one can use
lattice shaking to probe coherent band coupling and realize the
formation of ferromagnetic domains. Moreover, the dynamics induced
by shaking an optical lattice can lead to an admixture of excited
orbitals \cite{Strater} and constitutes an emergent branch of modern
quantum physics.

A substantial part of the previous studies has been primarily focussing on 
the renormalization of the physics due to driving, the mean-field approach \cite{Morsch} for weak
interactions, where the Gross-Pitaevskii equation is still valid, 
and a linear response treatment \cite{Iucci}. However, a
relatively large modulation of the strength or of the frequency of the
driving as well as strong interactions, calls for alternative
methods which can take into account higher-orbitals. Indeed, the
inclusion of higher-band contributions introduces new degrees of
freedom and as a result additional physical processes come into
play. Hereby, a sinusoidal shaking of the optical lattice is a natural starting 
point which induces an in-phase dipole mode on each site. An interesting and so far largely unexplored direction is the
study of the interplay between higher bands for the intra-well mode
and the inter-well tunneling dynamics with respect to the driving
frequency, and the investigation of the effect of the interatomic
interactions in the overall process. In this way, it is natural to
start with the investigation of the few body analogue in order to achieve a more comprehensive understanding of 
the microscopic properties of the strongly driven interacting system. Although the major part of the presented results is devoted to 
the case of four bosons in a triple-well setup, we provide strong evidence that our findings are still applicable for larger lattice systems 
and larger particle numbers.

Motivated by the recent experimental
progress \cite{Struck,Parker} we investigate in the present work the effects a periodically driven one-dimensional
optical lattice can introduce in a small ensemble of ultracold bosons. The
dynamical response of the system for a wide range of driving
frequencies is studied by means of the concept of fidelity 
or autocorrelation function. Even
though we consider a scenario with a deep lattice such that the
tunneling modes have a minor influence on the overall dynamics, a quite rich excitation
spectrum is found. We note that such intra-band excitations, which lead to a coupling between the two 
lowest energy bands, have been exploited in order to realise single- and two-qubit gates, where the 
quantum bit has been encoded in the localised Wannier functions of the two lowest energy bands 
of each lattice site \cite{Schneider}. In order to analyze the
intra-well dynamics we employ the one-body reduced density matrix. The Fourier spectrum of the local one-body
density as well as of the on-site density oscillations are employed in order to
obtain insights into the excited intra-well modes. We
find a resonant behaviour of the dipole mode indicating that the
intra-well dynamics can be controlled by adjusting the driving
frequency. Moreover, the magnification of the intra-well generated mode
at resonance is also manifested in the population of additional
lattice momenta. Our investigation of the resonances
is supported by a Floquet analysis for the
effective single-particle degree of freedom. This allows us to further explore 
the on-site dynamics and the inter-well tunneling
that occur due to the driving. Including interatomic interactions for larger 
atom numbers we analyze similarities and differences with
respect to the single-particle description. The above outlined
findings are confirmed for different filling factors, lattice
potentials, and boundary conditions. To solve the underlying 
many-body Schr\"{o}dinger equation we apply the $ab-initio$ MultiConfiguration
Time-Dependent Hartree method for Bosons (MCTDHB) \cite{Alon,Alon1}
which is especially designed to treat the driven 
out-of-equilibrium quantum dynamics of interacting bosons.

This article is organized as follows. In Sec.II we introduce our
setup and the multi-band expansion. Sec. III contains the
driven quantum dynamics first from a single-particle perspective, by performing a Floquet
analysis, and second by inspecting the dynamics of a small bosonic ensemble including repulsive interactions.
We summarize our findings and provide an outlook in Sec.V. Appendix A briefly outlines 
our computational method.

\section{Hamiltonian and multi-band expansion}
This section is devoted to a brief presentation of the
theoretical framework of our study. In particular, we shall briefly discuss 
the driven optical lattice, the underlying many-body Hamiltonian, and the concept of multi-band expansion. 
The latter will be a useful tool
in order to understand the excitations involved in the dynamics.

\subsection{Modeling the periodically-driven potential}

The periodic driving of an optical lattice can be accomplished in two
different ways. Retroreflecting mirrors that are used to form the lattice can be moved
periodically in space or, alternatively, a frequency
difference between counterpropagating laser beams can be induced by
means of acousto-optical modulators \cite{Parker} which renders the lattice 
time-dependent. Here, we model the driven optical lattice with  
a sinusoidal function of the form
\begin{equation}
\label{eq:1}{V_{sh}}({x},t) = {V_0}{\sin
^2}[k_{0}({x}-A\sin\omega_{D}{t})].
\end{equation}
Such a potential has been implemented in the experiment of e.g. ref. \cite{Gemelke}.
It is characterized by the barrier depth ${V_0}$, a lattice
wave-vector $k_{0} = \frac{\pi }{l}$, where $l$ denotes the distance
between successive potential minima, the amplitude $A$ and the
frequency $\omega_{D}=2\pi/T_{D}$ of the driving field. In an 
experiment $k_{0}$ is the wave vector of the laser beams which form the
optical lattice, while its depth ${V_0}$ can be tuned by adjusting
the lasers intensity.

\subsection{The Hamiltonian}

The Hamiltonian of $N$ identical ultracold bosons of mass $M$
confined in a driven one-dimensional $m$-well optical lattice reads
\begin{equation}
\label{eq:2}H = \sum\limits_{i = 1}^N { - \frac{{{\hbar ^2}}}{{2M}}}
\frac{{{\partial ^2}}}{{\partial x_i^2}} + {V_{sh}}({x_i},t) +
\sum\limits_{i < j} {{V_{{\mathop{\rm int}} }}({x_i} - {x_j})},
\end{equation}
where ${V_{{\mathop{\rm int}} }}({x_i} - {x_j}) = {g_{1D}}\delta
({x_i} - {x_j})$ denotes the short-range contact interaction
potential between particles located at position ${x_i}$, $i =
1,2,...,N$. In the ultracold regime the interaction is well
described by s-wave scattering whose 
effective 1D coupling strength \cite{Olshanii} is given by ${g_{1D}} =
\frac{{2{\hbar ^2}{a_0}}}{{Ma_ \bot ^2}}{\left( {1 - \frac{{\left|
{\zeta (1/2)} \right|{a_0}}}{{\sqrt 2 {a_ \bot }}}} \right)^{ -
1}}$. Here ${a_ \bot } = \sqrt
{\frac{\hbar }{{M{\omega _ \bot }}}}$ is the transverse harmonic oscillator length with ${{\omega _ \bot }}$ the
frequency of the two-dimensional confinement, while ${a_0}$ denotes the free space 3D s-wave
scattering length. In this way, the interaction strength can be
tuned either via ${a_0}$ with the aid of Feshbach resonances
\cite{Kohler,Chin}, or via the transversal confinement frequency
${\omega _ \bot }$ \cite{Kim,Giannakeas}.

For the sake of simplicity and computational convenience, we rescale
the Hamiltonian (2) in units of the recoil energy ${E_R} =
\frac{{{\hbar ^2k_{0}^2}}}{{2M}}$. Then, the corresponding length, 
time and frequency scales are given in units of ${k_{0}^{ - 1}}$, 
$\omega_{R}^{-1}=\hbar E_R^{ - 1}$ and $\omega_{R}$ respectively. In our simulations
we have used a sufficiently large lattice depth with values ranging
from $V_{0}=4.5$ $E_{R}$ to $8.0$ $E_{R}$ such that each well
includes three localized single-particle Wannier states. In particular, 
due to the deep optical lattice and small driving amplitudes (in comparison to the
lattice constant) mainly used in our simulations highly energetic excitations above
the barrier are excluded and as a consequence heating processes can
be minimized. The confinement of the bosons in the $m$-well system is imposed by the
use of hard-wall boundary conditions at positions $x_{\sigma} = \pm
\frac{{m\pi }}{{2k_{0}}}$, where the potential is maximum. In
addition, we set also $\hbar = M = k_{0} = 1$ and the coupling
strength becomes $g = \frac{{{g_{1D}}}}{{{E_{R}}}}$, while $A$
represents the dimensionless driving amplitude. The rescaled
shaken triple well is given by ${V_{sh}}({x_i},t) = {V_0}{\sin
^2}({x_i}-A\sin\omega_{D}{t})$ with the hard wall boundaries located
at $x_{\sigma}= \pm 3\pi /2$.

\subsection{The multi-band expansion}

The understanding of the spatial localization of states in lattice systems makes the use
of multi-band Wannier number states crucial as it includes the
information of excited bands and allows to interpret both intraband and
interband processes. In general, this representation is
valid when the lattice potential is deep enough such that the
Wannier states between different wells have a very small overlap for
not too high energetic excitation. In the present case where the
potential is periodically driven the above description can still be used as long
as the driving amplitude is small enough in comparison to the
lattice constant $l$, i.e. $A\ll l$. In this way, each localized Wannier
function can be still adapted and assigned to a certain well and the respective
band-mixing is fairly small. For large displacements one should use
a time-dependent Wannier basis in order to ensure that the corresponding
on-site Wannier states are well-adapted to each well during the driving.

To introduce the formalism, let us consider a system
consisting of $N$ bosons, $m$-wells and $k$ localized single
particle bands \cite{Mistakidis,Mistakidis1}. Then, the expansion of the many-body 
bosonic wavefunction in terms of the number states of non-interacting bosons reads
\begin{equation}
\label{eq:3}\left| \Psi  \right\rangle  = \sum\limits_{\{N_{i}\},\textbf{I}}
{{C_{\{N_i\};\textbf{I}}}{{\left| {{N_1},{N_2},...,{N_m}} \right\rangle
}_\textbf{I}}},
\end{equation}
where ${{\left| {{N_1},{N_2},...,{N_m}} \right\rangle
}_\textbf{I}}$ is the multiband Wannier number state and the element
$N_{i}$ denotes the number of bosons being localized in the
$i$-th well satisfying the constraint
$\sum_{i=1}^{m}N_{i}=N$. The summation is performed over the
different configurations of the $N$ bosons according to their
energetical order denoted by the index $\textbf{I}$. In particular,
the index $\textbf{I}$ corresponds to a high dimensional quantity
$\textbf{I}=(\textbf{$I_{1}$},\textbf{$I_{2}$}$,...,$\textbf{$I_{m}$}$)
which contains $m$ elements each of them being a $k$-component
vector. More precisely, the $q$-th element can be written as
$\textbf{$I_{q}$}$=($\textbf{$I_{q}^{(1)}$}$,
$\textbf{$I_{q}^{(2)}$}$,..., $\textbf{$I_{q}^{(k)}$}$), where
$\textbf{$I_{q}^{(k)}$}$ refers to the number of bosons located at
the $q$-th well and $k$-th band, satisfying the constraint
$\sum_{q=1}^m\sum_{i=1}^k I_q^{(i)} = N$.
Within the above notation one can investigate, among others, the
probability of $N_{0}<N$ bosons to be in an excited band or to
find a specific number state configuration. Indeed, suppose the case
of $N_{0}<N$ bosons excited in the $i$-th band while the rest
$N-N_{0}$ lie in lower bands. Then, it must hold
$\textbf{$I_{1}^{(j)}$}=\textbf{$I_{2}^{(j)}$}=...=\textbf{$I_{m}^{(j)}$}=0$
for every $j>i$, while
$\textbf{$I_{1}^{(i)}$}+\textbf{$I_{2}^{(i)}$}+...+\textbf{$I_{m}^{(i)}$}=N_{0}$
and $\textbf{$I_{1}^{(1)}$}+...+\textbf{$I_{m}^{(j_{1})}$}=N-N_{0}$
for every $j_{1}<i$.

Let us consider an example of a system with four bosons ($N=4$) confined in a triple well
($m=3$) which includes three bands ($k=3$). Then, for instance, the state
${\left| {1,2,1} \right\rangle_\textbf{I}}$ with
$\textbf{I}=(\textbf{$I_{L}$},\textbf{$I_{M}$},\textbf{$I_{R}$})$,
and $\textbf{$I_{L}$}$=$\textbf{$I_{R}$}$=(0,1,0),
$\textbf{$I_{M}$}$=(0,1,1) denotes a state for which in the left (right) well
one boson occupies the first excited band, whereas in the middle well
one boson is localized in the first excited and one in the second
excited band. As a final attempt, here, we make a link between the ground state and its dominant spatial 
configuration in terms of the aforementioned multiband expansion. To do that, let us choose again 
a system consisting of four bosons in a triple well as it will extensively be used in the following. It is known that, 
in general, the ground state configuration depends on the interaction strength, while for the present system, i.e. $N=4$ and $m=3$, the on-site interaction 
effects will always be prominent. For the non-interacting case ($g=0$) the dominant spatial configuration of the 
system is ${\left| {1,2,1} \right\rangle_\textbf{I}}$, with $I_{L}=I_{R}=(1,0,0)$ and $I_{M}=(2,0,0)$ due to the hard-wall boundaries which render the middle and outer sites non-equivalent. 
In the course of increasing interaction a tendency 
towards a uniform population of each site, e.g. for $g=0.2$, due to the repulsion of the bosons is observed. In this region the system is described by a superposition of lowest-band states 
which are predominantly of single-pair occupancy, e.g. ${\left| {1,2,1} \right\rangle_\textbf{I}}$, ${\left| {2,1,1} \right\rangle_\textbf{I}}$, and double-pair occupancy, e.g. ${\left| {2,2,0} \right\rangle_\textbf{I}}$. 
For further increasing repulsion, e.g. $g=0.4$, a trend towards 
the repopulation of the central well is noted. As we enter the strong interaction regime, e.g. $g=1.5$, the state consists of a particle in the first excited-band being 
on a commensurate background of localized particles which lie in the zeroth band and the dominant ground state configuration is ${\left| {1,2,1} \right\rangle_\textbf{I}}$, with $I_{L}=I_{R}=(1,0,0)$ and $I_{M}=(1,1,0)$. Finally, for 
strong interparticle repulsion, e.g. $g=3$, the contribution from the higher-band states becomes more prominent and the corresponding ground state configuration is characterized by an admixture of zeroth- and excited-band states.
\begin{figure}[ht]
        \centering
           \includegraphics[width=0.50\textwidth]{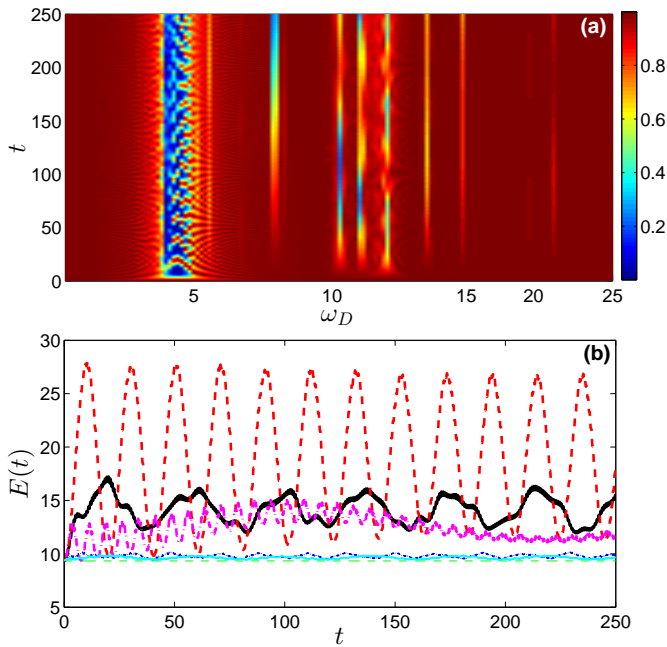}
                \caption{(a) Time evolution of the fidelity $F_{\omega_{D}}(t)$ as a function of the driving frequency $\omega_{D}$ (measured in units of $\omega_{R}$). (b) Time evolution of the expectation value of the Hamiltonian (2) (measured in units of the recoil energy $E_{R}$)
                for various driving frequencies $\omega_{D}=0.4$ (green thin dashed line), $\omega_{D}=4.0$ (black thick solid line), $\omega_{D}=4.5$ (red thick dashed line),
                $\omega_{D}=5.25$ (magenta thick dashed-dotted line), $\omega_{D}=11.0$ (blue thin dashed-dotted line), and $\omega_{D}=13.375$ (light-blue thin solid line).
                The driving amplitude is $A=0.05$, while the initial state corresponds to the ground state of four
                weakly interacting bosons with $g=0.1$ confined in a triple-well. Time unit is $\omega_{R}^{-1}$.}
\end{figure}

\section{Driven quantum dynamics}
This section is devoted to a detailed analysis of the bosonic 
dynamics in a driven optical lattice. At the beginning,  
a general overview of the effect of the driving on the finite bosonic
ensemble with respect to the driving frequency is given.
Subsequently, a Floquet analysis is employed in order to
investigate the underlying single-particle physics. Finally, we focus on specific interaction 
effects.

\subsection{Dynamical response}

Let us explore the dynamical response or sensitivity of the
system with respect to the driving frequency $\omega_{D}$. In order
to investigate the stability of the system against the perturbations
induced by the shaking [see Eq.(1)], we first analyse the fidelity
\cite{Gorin} between the initial state and the state evolved at time
$t$: $F_{\omega_{D}}(t)=| \left\langle {\Psi (0)} \right|\left.
{\Psi (t)} \right\rangle|^2 $, where the dependence on $\omega_D$ is
implicit in the time evolved state $\Psi(t)$. Here we will
consider a system of four bosons in a triple-well with
$g=0.1$, whose ground state (i.e., the initial state $\Psi(0)$)
corresponds to a superfluid state, as the filling factor is not
commensurable and we do not encounter the formation of a Mott insulating
state. In terms of its dominant spatial configuration our system initially consists (see also Sec.II.C) 
of two bosons in the middle well and two others each of them localized in one of 
the outer wells, i.e. the state ${\left| {1,2,1} \right\rangle_\textbf{I}}$, 
with $I_{L}=I_{R}=(1,0,0)$ and $I_{M}=(2,0,0)$ has the most prominent contribution. Figure 1(a) shows $F_{\omega_{D}}(t)$ as a function of the
driving frequency $\omega_{D}$. The dynamics is characterized by
three main regions with respect to $\omega_{D}$, where the system is
driven far from the initial state, while for the remaining  
frequency regions (red sections in Fig. 1(a)) the evolved state is
essentially unperturbed by the driving. In the first region, between
$4.0<\omega_{D}<5.5$, the minimal overlap in the course of the dynamics drops
down to $0.1$, whereas in the second ($7.0<\omega_{D}<8.0$) and
third ($10.0<\omega_{D}<15.0$) regions the system maximally departs from the
initial state with a percentage on the order of $50\%$ and $65\%$,
respectively. The emergence of these dynamical regions strongly
depends on the parameters of the optical lattice. For instance, for
smaller lattice depths the aforementioned regions will be wider,
because of the smaller potential energy, which favors a possible
deviation of the system from the initial state.

Let us inspect the time evolution of the total energy $E(t)=
\left\langle \Psi(t) \right| \hat H(t)\left| \Psi(t) \right\rangle$.
Figure 1(b) shows $E(t)$ for various driving frequencies
$\omega_{D}$. For driving frequencies where $F_{\omega_{D}}\simeq1$
[e.g, $\omega_{D}\in\{1,3\}$, see also Fig. 1(a)] the dependence of
the energy on the driving frequency is weak and it is essentially
constant during the time evolution. On the other hand, for the
regions where $F_{\omega_{D}}\ll1$, $E(t)$ increases initially and
it shows an oscillatory behaviour. In particular, for
$\omega_{D}=4.5$ the total energy exhibits an oscillatory (almost periodic) pattern which
can also be observed in the corresponding fidelity evolution. This driving
frequency will be referred to in the following as critical and denoted by
$\omega_D^c$, that is, the driving frequency for which
$\min_{t\in [0,T]}F_{\omega_D}(t)$ is minimal. Indeed, as we shall see below,  
the most interesting dynamics of the system takes place
close to this frequency.
\begin{figure*}[ht]
        \centering
           \includegraphics[width=0.80\textwidth]{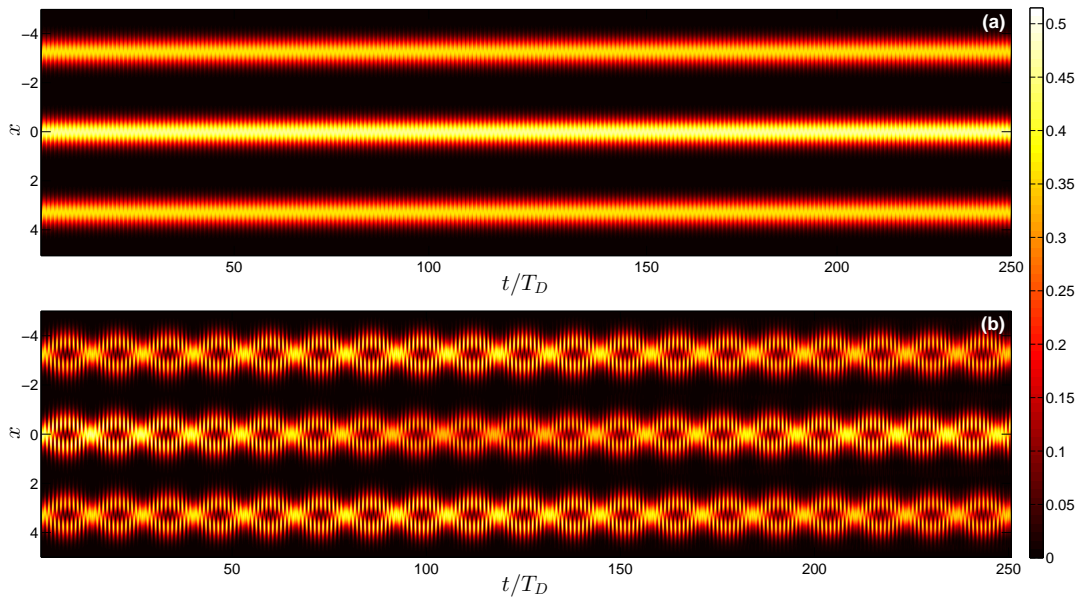}
                \caption{Time evolution of the one-body density $\rho_{1}(x,t)$ in a triple-well
                potential for different driving frequencies: (a) $\omega_{D}=2.0$ (top panel) and (b) $\omega_{D}=4.5$ (lower panel). The
                driving amplitude is fixed to the value $A=0.05$, while the initial state corresponds to the ground state of four
                weakly interacting bosons with $g=0.1$. The spatial extent of the lattice is expressed in units of $k_{0}^{-1}$, while the time units are rescaled in terms of the driving period $T_{D}$.}
\end{figure*}

Finally, let us inspect the response of the system to the
driving from a one-body perspective via the single-particle density
$\rho_{1}(x,t) = \int {d{x_2}...d{x_N}{{\left| {\Psi
(x,{x_2},...,{x_N};t)} \right|}^2}}$. Figure 2 illustrates the evolution of the one-body
density for different driving frequencies $\omega_{D}$, but with the
same amplitude $A$. The driving leads to oscillations of the
particles densities in every site. As it can be observed by having a glance at
Fig. 2(a), the one-body density shows a weak response for driving
frequencies away from the critical region $\omega_D\in [4,5.5]$,
while for $\omega_{D}=\omega_D^c$ [see Fig. 2(b)] we observe the
periodic formation of enhanced density oscillations being
accompanied by a broadening of each intra-well ensemble. 
The peculiar behaviour of the bosonic ensemble
observed for $\omega_{D}=\omega_D^c$ is characterized by three
processes and time scales: $i$) the internal fast oscillations of
the density; $ii$) the large amplitude oscillations of the density
in each well of period $\sim 14$; $iii$) the tunneling
between the wells with a period of about $200$. All these
features will be analyzed in detail in the following subsections
both at the single particle and many-body level.
\begin{figure}[ht]
        \centering
           \includegraphics[width=0.30\textwidth]{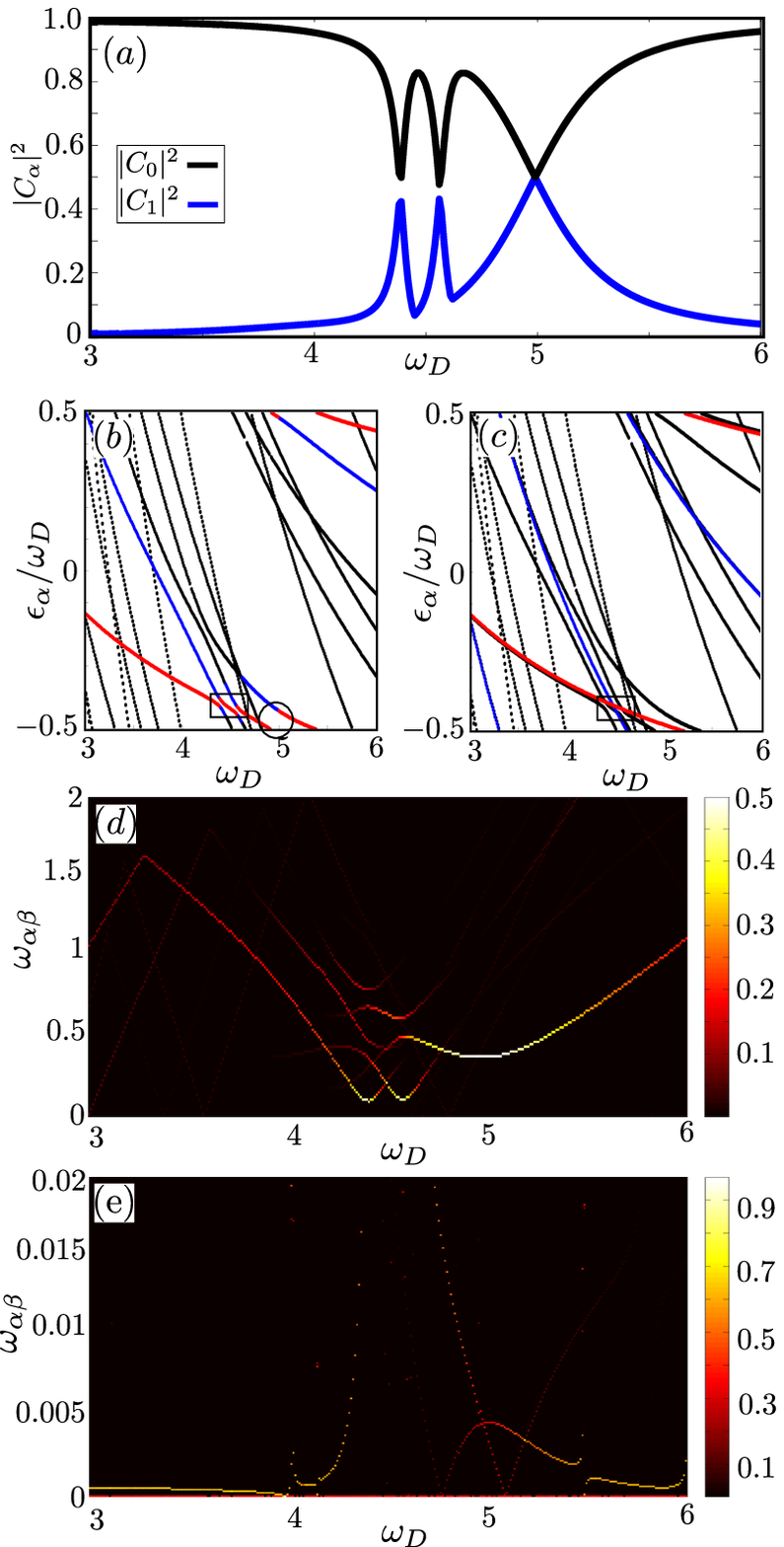}
                \caption{\label{fig3}On-site dynamics for a single particle. (a) Populations $|C_0|^2$ and $|C_1|^2$ of the two most populated FMs. (b)
QE spectrum as a function of the driving frequency $\omega_D$ (measured in units of $\omega_{R}$). Highlighted are the most (red) and second most (blue) populated FMs.
The rectangular area indicates the narrow avoided crossings, while the circle highlights the area where a broad avoided crossing among the FMs appears with respect to the driving frequency. 
(c) In black is again the QE spectrum (same as in (b)). Additionally, we show the most (red) and second most (blue) populated states of the static, i.e. undriven, lattice. 
For comparison, we depict again the black rectangle at the same position as in (b).
(d) Frequencies $\omega_{\alpha \beta}$ of the on-site dynamics as a function of the driving frequency (see main text).
(e) Same as (d), but in the triple well setup (shown is only the extract of small frequencies $\omega_{\alpha \beta} \ll \omega_D$ which corresponds to tunneling dynamics). In all panels $A=0.05$.}
\label{fig3}
\end{figure}
\begin{figure}[ht]
        \centering
           \includegraphics[width=0.50\textwidth]{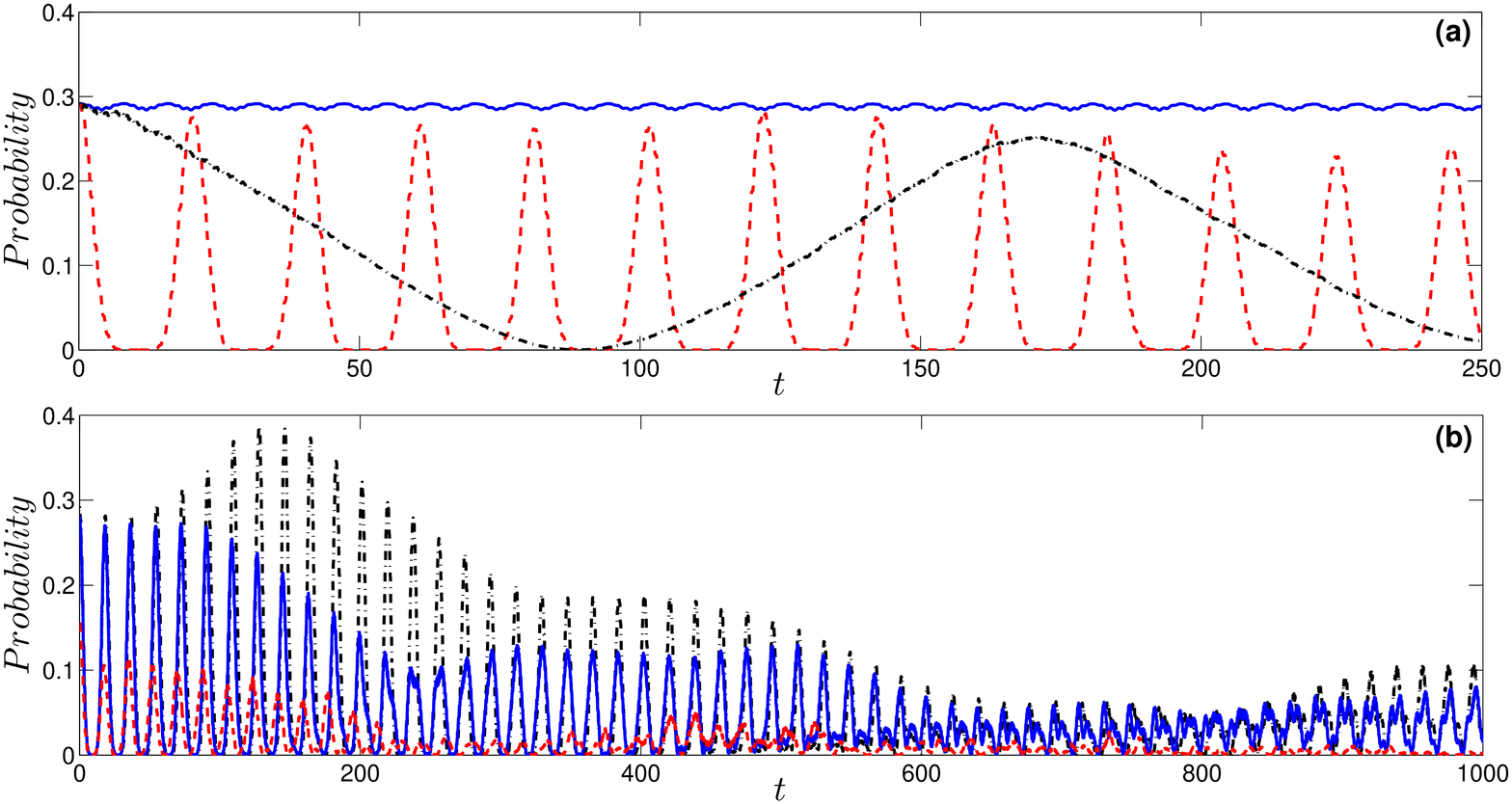}
                \caption{(a) Tunneling probability (see main text) $|D_{\{N_i\};\textbf{I}}|^2=|{}_{\textbf{I}}\left\langle 2,1,1| \Psi(t)
                 \right\rangle|^2$ with $I_{L}=(2,0,0)$ and $I_{R}=I_{M}=(1,0,0)$ as a function of time
                for different driving frequencies $\omega_{D}=0.5$ (blue solid line), $\omega_{D}=4.5$ (red dashed line) and $\omega_{D}=11.0$ (black dashed-dotted
                line). The most
                significant contribution of the interband tunneling mode is between the state ${\left| {2,1,1} \right\rangle _\textbf{I}}$ [with $I_{L}=(2,0,0)$ and $I_{R}=I_{M}=(1,0,0)$] and the initial ${\left| {1,2,1} \right\rangle _\textbf{I}}$
                [with $I_{M}=(2,0,0)$ and $I_{R}=I_{L}=(1,0,0)$].
                (b) Inter-well tunneling probability $|D_{\{N_i\};\textbf{I}}|^2$ at resonance for different values of the interatomic 
                interaction $g=0.1$ (black dashed-dotted line), $g=0.5$ (blue solid line) and $g=2.0$ (red dashed line). In all panels $A=0.05$. The 
                time evolution is expressed in units of $\omega_{R}^{-1}$.}
\end{figure}

\subsection{Single particle dynamics}
\label{S1}

Here we investigate to which extend the previously
presented results can be understood in the limit of zero interaction
among the particles by means of Floquet theory. Specifically, we are interested in two distinct
features of the dynamics observed in Fig. 2(b): First, the on-site
dynamics and, especially, its resonance-like dependence on the driving
frequency $\omega_D$, and second, the inter-well tunneling dynamics
which is enhanced at certain values of $\omega_D$.

\subsubsection{Floquet theory}
\label{S1.1} To be self-contained, we start by summing up the main
notions of Floquet theory. Because of the temporal periodicity of
the single particle Hamiltonian employed throughout this work [Eq. (2)
with $g=0$ and $N=1$], every solution of the time-dependent
Schr\"{o}dinger equation (TDSE) takes the form of a Floquet mode
(FM) $\Psi_{\alpha}(x,t)$ which in turn can be written as:
$\Psi_{\alpha}(x,t)=e^{-i\epsilon_{\alpha}t/\hbar}
\Phi_{\alpha}(x,t)$ with the real quasi energy (QE)
$\epsilon_{\alpha} \in [-\hbar\omega_D/2, +\hbar\omega_D/2]$ and
with $\Phi_{\alpha}(x,t) = \Phi_{\alpha}(x,t+T_D)$ respecting the
temporal periodicity of the Hamiltonian \cite{Tannor}. 
The FMs are eigenvectors of the time evolution operator over one driving period
 \begin{equation}
 U(T_{D}+t_0,t_0)\Psi_{\alpha}(x,t_0)=e^{-i\epsilon_{\alpha} T_D/\hbar}\Psi_{\alpha}(x,t_0).
 \label{EV}
\end{equation}
This property is of particular interest as it allows for a stroboscopic time evolution of an arbitrary initial state $\Psi(x,t_0)$ once the FMs of a system are know.
To show this, we exploit the fact that the FMs constitute an orthonormal basis for the solution space of the TDSE \cite{Hanggi} and expand $\Psi(x,t_0)$ at the initial time $t=t_0$ as
 \begin{equation}
  \Psi(x,t_0)= \sum_{\alpha} C_{\alpha}(t_0) \Psi_{\alpha}(x,t_0)
 \label{expand}
\end{equation}
 with the corresponding coefficients $C_{\alpha}(t_0)$. By applying the one period evolution operator
$U(T_{D}+t_0,t_0)$ on both sides of Eq. (5) for $m$ times
and by virtue of Eq. (4),
we readily obtain the stroboscopic time evolution of $\Psi(x,t_0)$ as
\begin{equation}
 \Psi(x,t_0+mT)= \sum_{\alpha} C_{\alpha}(t_0) e^{-i\epsilon_{\alpha}mT_D/\hbar } \Psi_{\alpha}(x,t_0).
 \label{time evo}
\end{equation}
Numerically, we obtain the FMs
for a given initial time $t_0$ by calculating the eigenvectors of the one period evolution operator $U(T_{D}+t_0,t_0)$ [see Eq. (4)]. 
We refer the interested reader to ref. \cite{Wulf:2014} for a detailed description of the employed computational scheme.

Finally, let us note that Eq. (6) already reveals some interesting features of the time evolution in periodically driven systems as we shall see in the following.
Imaging that only a single FM, say $\Psi_0(x,t)$, is populated. The stroboscopic  evolution of the probability density is thus given as $|\Psi(x,mT_D)|^2 = |C_{0}|^2 |\Psi_{\alpha}(x,0)|^2$.
Hence, $|\Psi(x,t)|^2$ is again periodic with period $T_D$ and the only time dependence arises from the explicit time dependence of the FM, which is commonly referred to as ''micro-motion''.
This situation changes if the initial state populates multiple FMs. In this case one encounters interference terms in $|\Psi(x,mT_D)|^2$ between the different FMs in the form of
$\sim e^{im(\epsilon_{\alpha}-\epsilon_{\beta})T_D/\hbar}$.  Thus, the quasi energies $\epsilon_{\alpha}$ in periodically driven systems play a comparable role in the time evolution as the energy eigenvalues do in
time independent setups.
\begin{figure}[ht]
        \centering
           \includegraphics[width=0.50\textwidth]{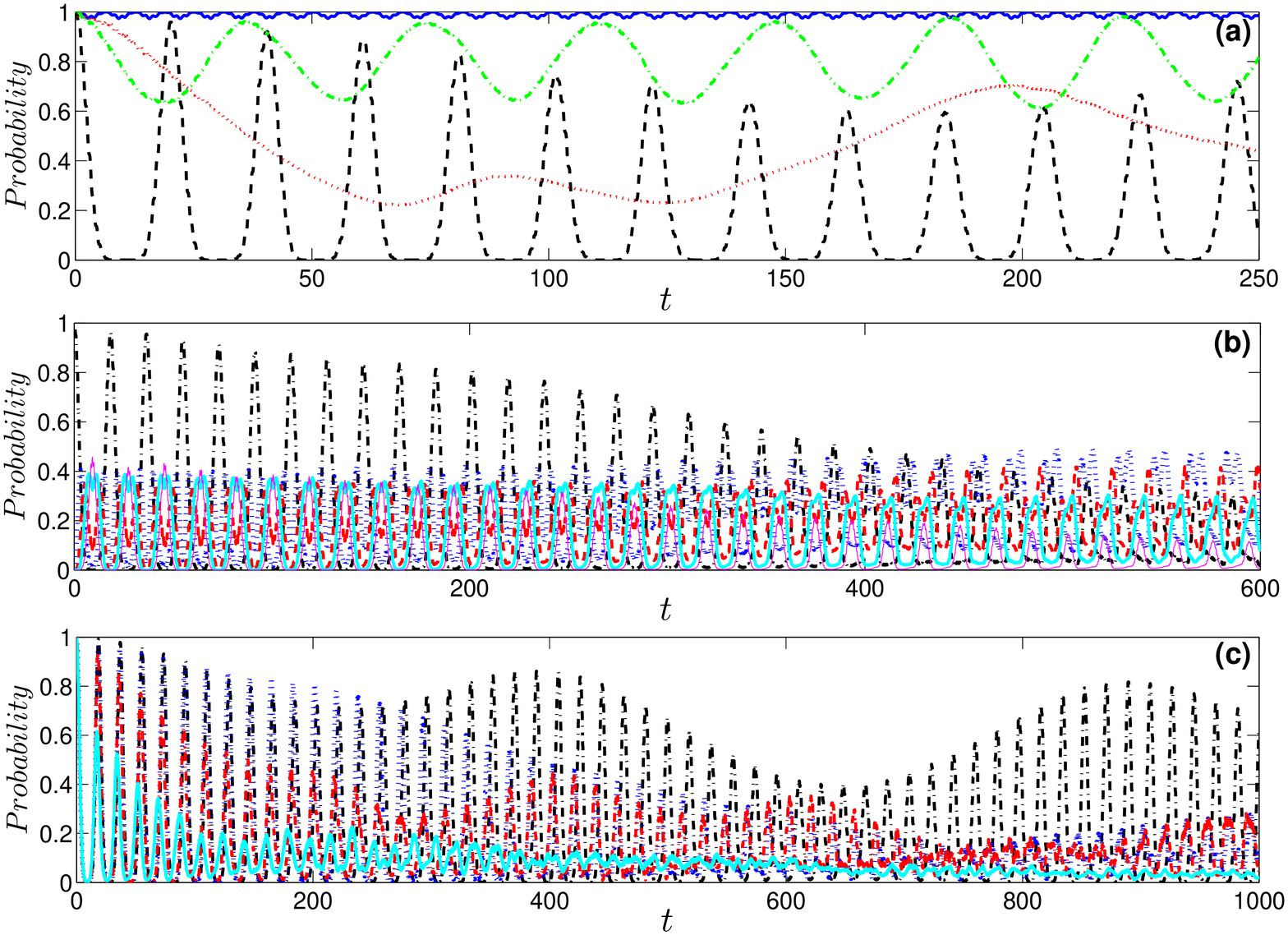}
                \caption{(a) Probability $|B_{\{N_i\};\textbf{I}}|^2$ (see main text) for all bosons to be in the zeroth-band during the evolution for different driving
                frequencies $\omega_{D}=2.0$ (blue solid line), $\omega_{D}=4.5$ (black dashed line), $\omega_{D}=10.25$ (red dotted line) and $\omega_{D}=11.75$ (green dashed-dotted line). (b) Comparison of different
                 excitation scenarios at $\omega_{D}=4.375$. The black dashed-dotted line refers to the probability for 
                 all the bosons to be in the zeroth-band while the blue dotted, red dashed, light-blue thick solid and magenta thin solid line refer to the probability to have
                 one, two, three or four bosons, respectively, in the first-excited band. (c) Probability $|B_{\{N_i\};\textbf{I}}|^2$ for all the bosons to be in the
                 zeroth-band for $\omega_{D}=\omega_D^c$, but with different interparticle repulsion $g=0$ (black dashed-dotted), $g=0.1$ (blue dotted line), $g=0.5$ (red dashed-dotted line)
                 and $g=2.0$ (light-blue solid line). In all panels $A = 0.05$. The time evolution is expressed in units of $\omega_{R}^{-1}$. }
\end{figure}

\subsubsection{On-site dynamics in the single well}
\label{S1.2}

To begin with, we shall investigate the observed on-site dynamics
[see Fig. 2(b)], and in particular their dependence on the driving
frequency $\omega_D$. To this end, we simplify the setup studied
in Sec. III.A to just a single well of the lattice
potential. Hence, the potential is given by $V_{sh}(x,t)=V_0 \sin^2
(x-A\sin(\omega_D t))$ for $x \in (-\pi,+\pi]$ and we
impose periodic boundary conditions at $x=\pm \pi$ in order to
mimic the situation in an extended lattice. We choose as 
initial state $\Psi(x,0)$ the single particle density as shown in
Fig. 2(b) at $t=0$ within the central potential well. The time
evolution is then obtained by expanding $\Psi(x,0)$ in terms of the
FMs of the system and by making use of Eq. (6). As a
result, we find that we can reproduce some of the main features of
the on-site dynamics shown in Figs. 2(a) and (b), namely, we observe 
resonantly enhanced on-site oscillations in an interval of the
driving frequencies around $\omega_D \approx 4.5$. Following the
discussion in ref. \cite{Wulf:2015}, further insight into this
effect can be obtained by studying the population of the FMs by the
initial state as a function of $\omega_D$. We therefore sort the FMs
$\Psi_{\alpha}$ according to their overlap with the initial state
and label the mode with the largest overlap as $\Psi_{0}$, the mode
with the second largest overlap as $\Psi_{1}$ etc. In Fig. 3(a)
the coefficients of the two most populated
FMs, $|C_0|^2$ and $|C_1|^2$, are shown as a function of the driving
frequency. Apparently, both at small frequencies ($\omega_D \lesssim
4$) and at large ones ($\omega_D \gtrsim 5.5$) only a single FM is
notably populated, while $|C_0|^2$ and $|C_1|^2$ become comparable
at distinct driving frequencies (e.g., at $\omega_D \approx 5$).
According to our discussion above, in cases when $|C_0|^2$ is close
to one, and thus only a single FM is populated, the stroboscopic
time evolution, as given by Eq. (6), becomes, to a good
approximation, time periodic with the period of the driving $T_D$.
Note that this agrees with the observation of Fig. 2(a), that away
from the resonance frequencies, the single particle density merely
performs oscillations whose period matches $T_D$. This corresponds precisely to  
the previously described micro-motion arising from the explicit time
dependence of the FM $\Psi_{0}(x,t)$.

On resonance, when $|C_0|^2 \approx |C_1|^2 $, the evolution of
$|\Psi(x,0)|^2$ includes, besides the micro-motion, an interference
term between $\Psi_0$ and $\Psi_1$, whose period is dictated by the
corresponding quasi energies and is given by:
$T_{\text{osc}}/T_D=\hbar \omega_D / (\epsilon_1-\epsilon_0)$.
Indeed, we find that this term is responsible for the observed
on-site mode with a period of $\sim 14$ lattice oscillations
[compare Fig. 2(b)]. Up to now, however, it is not yet clear why
$\Psi_1$ is resonantly populated at certain frequencies. In order to
provide an answer to this question we follow the argumentation in
ref. \cite{Wulf:2015} and consider the dependence of the quasi
energy spectrum on the driving frequency $\omega_D$ as shown in Fig. 3(b).
Highlighted are the two most populated modes at each
$\omega_D$ (blue and red dots) revealing avoided crossings of these
two modes at the frequencies where a resonant enhancement of
$|C_1|^2 $ was observed in Fig. 2(b). Hence, at these values of
$\omega_D$ the FMs $\Psi_0$ and $\Psi_1$ are resonantly coupled by
the driving which results in an increase of $|C_1|^2 $ and
ultimately to the in Sec. III.A described on-site dynamics.

In the following we provide insight into the question why we observe Floquet resonances at driving frequencies around $\omega_D \sim 4.5$. 
Let us start by noting that, by means of appropriate unitary transformations, the single particle Hamiltonian with a potential as given in Eq. (1) can be recasted into the form:
\begin{equation}
 \tilde H = -\frac{\hbar^2}{2m} \frac{\partial^2}{\partial x^2} + V_0 \sin^2(k_0 x) + V_D sin(\omega_D t) x,
\end{equation}
where the amplitude of the oscillating term is given by $V_D= m A \omega_D^2 $. That is the transformed Hamiltonian takes the form of a static lattice plus a time-dependent perturbation whose 
strength is determinded by $V_D$. For the used parameters of $m=1$ and $A=0.05$ and for the 
range of considered frequencies of $3 \lesssim \omega_D \lesssim 6$ we get that the amplitude $V_D$ of the time-dependent term is of order one. Hence, it can be seen as a small perturbation 
compared to the static term of strength $V_0=15$ and we can expect that the QEs of the driven lattice setup can be estimated by the actual energies of the undriven lattice. Resonances would than be expected whenever
the energy difference between two notably populated eigenstates of the static system matches an integer multiple of $\hbar \omega_D$.
In fact we find that the energies of the three energetically lowest states of even parity are given by $E_0\approx 2.6$, $E_1\approx 11.6$ and $E_2=16.9$. Naivly, 
we expect driving induced resonances whenever the ground state is resonantly coupled to one of the excited states. Indeed we find $E_1-E_0\approx 2\times 4.5$ and $E_2-E_0\approx 3 \times 4.8 $.
Thus, following this line of arguments, at driving frequencies of approximately $\omega_D=4.5$ the ground state of the unperturbed lattice is coupled via a 2 (3) photon process to the first (second) excited states. 
In order to justify this simplyfied picture we show the energies $E_0$ and $E_1$ on top of the QE spectrum of the driven lattice (see Fig. 3(c)). Away from any resonances, both energies are almost identical to
the QEs of the corresponding Floquet states, so e.g. the red line is practically on top of an underlying black line. Closer to the resonance region, we see of course deviations of the QEs from the mere energies 
of the undriven lattice as the different states are coupled by the driving.

Finally, Fig. 3(c) provides an overview over the possibly
observed frequencies in the on-site dynamics at various 
driving frequencies. Shown are the frequencies associated to all
possible interference terms between the FMs weighted by their
overlap with the initial state. More precisely, we calculate
$\omega_{\alpha \beta}=(\epsilon_{\alpha}-\epsilon_{\beta})/\hbar$
for all pairs of FMs at a given driving frequency and determine the
colour coding by computing the product $|C^*_{\alpha} C_{\beta}|$.
Hence, the frequency $\omega_{\alpha \beta}$ appears in Fig. 3(c)
only when both of the corresponding FMs
$\Psi_{\alpha}$ and $\Psi_{\beta}$ have appreciable overlap with the
initial state. In agreement with the discussion concerning Fig. 3(b)
we observe pronounced on-site oscillations only
within an interval of driving frequencies $4.0 \lesssim \omega_D
\lesssim 5.5$. In particular, the two narrow avoided crossings
around $\omega_D^c \approx 4.5$ [see the rectangular in Fig. 3(b)] yield a low
frequency on-site dynamics, whereas the comparably broad avoided
crossing at $\omega_D\approx 5$ [see the circle in Fig. 3(b)] results in a much faster on-site
oscillation.

\subsubsection{Tunneling dynamics in the triple well}
\label{S1.2}

Besides the on-site dynamics, Fig. 2(b) revealed a pronounced
tunneling between the lattice sites at certain driving frequencies.
Similar to the previous section, we analyze this effect in the
following by applying Floquet theory for the single particle
dynamics. We choose the same setup as before, that is, 
$V_{sh}(x,t)=V_0 \sin^2 (x-A\sin(\omega_D t))$, with the same
initial state, i.e. essentially a Gaussian centered around the
potential well at $x=0$, but with the difference that the periodic
boundary conditions are imposed at $x=\pm 2\pi$ (instead of at
$x=\pm \pi$ as we did before). In this way we allow for
tunneling of the wave packet into the two neighbouring lattice
sites. As for the on-site dynamics, we provide an overview over the 
observable frequencies in the temporal evolution in Fig. 3(d)
[in close analogy to Fig. 3(c)]. Note that,
since the tunneling dynamics observed in Sec. III.A occurs on
much longer timescales as compared to the on-site dynamics, we only
show the extract of the regime of small frequencies, i.e.
$\omega_{\alpha \beta} \ll \omega_D$. Furthermore, because no
on-site dynamics occurs with timescales matching the extremely small
frequencies of $\lesssim 0.02$, all the frequencies depicted in Fig. 3(d)
are indeed associated with an inter-well tunneling
mode. In accordance with the observation made in the many-particle
simulations [cf. Fig. 2(b)] we observe a strong increase of the
frequencies associated with the tunneling dynamics in the range of
driving frequencies of $4\lesssim \omega_D \lesssim 5.5$. Away from
this resonance, for example at $\omega_D=2.5$, the only notable
tunneling mode corresponds to an interference term of two FMs which
oscillates with a period of $T_{\text{osc}}/T_D \approx 3300$ and
could therefore not be observed in the simulations performed in
Sec. III.A. Within the regime of resonant driving, e.g. at
$\omega_D=4.5$, the frequency of the tunneling mode is increased
strongly and the associated oscillation period becomes
$T_{\text{osc}}/T_D \approx 200$ matching the observed tunneling
mode in the weakly interacting regime [cf. Fig. 2(b)].

\subsection{Interband tunneling and excitation processes}

In the previous section we have shown that most of the features of the 
(effective) single-particle dynamics of Fig. 2 can be explained 
via a non-interacting Floquet theory. As we shall see now, however, the full dynamics
presents a rich excitation spectrum ascribable to the particles
interaction, especially in the strong interaction regime. Thus, we
investigate the tunneling and excitation probabilities of the
dominant particle configurations, for different driving frequencies
$\omega_{D}$, by means of the multiband expansion introduced in
Sec.~II.C. More precisely, we compute and analyze the
probabilities, during the dynamics, defined as
\begin{equation}
\label{eq:7}|C_{\{N_i\};\textbf{I}}|^2=|{}_{\textbf{I}}\left\langle
N_{1},N_{2},N_{3}| \Psi(t) \right\rangle|^2.
\end{equation}
The case $I_{m}^{(k)}=0$ $\forall$ $k>1$ refers to the lowest-band inter-well tunneling
dynamics. The initial state of the system corresponds to the ground state of four weakly 
interacting bosons with $g=0.1$ in a triple well, while the dominant number state configuration (see also Sec.II.C) is 
${\left| {1,2,1} \right\rangle_\textbf{I}}$ with $I_{L}=I_{R}=(1,0,0)$ and $I_{M}=(2,0,0)$. In this way, a lowest-band tunneling process 
can take place among the initial state and: a) another state of single-pair occupancy, 
e.g. ${\left| {2,1,1} \right\rangle_\textbf{I}}$ ($I_{L}=(2,0,0)$ and $I_{M}=I_{R}=(1,0,0)$), b) a state with double-pair occupancy,  
e.g. ${\left| {2,2,0} \right\rangle_\textbf{I}}$ ($I_{L}=I_{M}=(2,0,0)$ and $I_{R}=(0,0,0)$), c) a state with triple occupancy, 
e.g. ${\left| {3,1,0} \right\rangle_\textbf{I}}$ ($I_{L}=(3,0,0)$, $I_{M}=(1,0,0)$, $I_{R}=(0,0,0)$) or d) a state with quartic occupancy, 
e.g. ${\left| {4,0,0} \right\rangle_\textbf{I}}$ ($I_{L}=(4,0,0)$ and $I_{M}=I_{R}=(0,0,0)$). However, the from the system prefered tunneling processes 
form a hierarchy according to the energetical difference between the initial and final state. For instance, a tunneling process to another state of single-pair 
occupancy will be more preferable than to a state of double-pair occupancy etc. Figure 4(a) 
shows the tunneling probability to the energetically closest number state, which is ${\left| {2,1,1} \right\rangle_\textbf{I}}$ (or ${\left| {1,1,2} \right\rangle_\textbf{I}}$) with $I_{L}=(2,0,0)$ and $I_{M}=I_{R}=(1,0,0)$ 
(or $I_{L}=I_{M}=(1,0,0)$ and $I_{R}=(2,0,0)$), i.e. $|D_{\{N_i\};\textbf{I}}|^2=|{}_{\textbf{I}}\left\langle 2,1,1| \Psi(t)
\right\rangle|^2$ with $I_{L}=(2,0,0)$ and $I_{M}=I_{R}=(1,0,0)$,
for various driving frequencies. As it is shown, for
$\omega_{D}<\omega_D^c$ this tunneling mode has a small amplitude
and it is quite insensitive to $\omega_{D}$ as intuitively expected from the fact that the evolved-state is essentially unperturbed by the driving (see also Fig. 1(a)). For
$\omega_{D}\approx\omega_D^c$, however, the amplitude of the oscillations
is significantly larger indicating an enhancement of the tunneling (see also Figs. 2(b) and 3(d)), whereas when $F_{\omega_{D}}\ll1$ and
$\omega_{D}>\omega_D^c$ ($\omega_{D}=11.0$ curve in Fig. 4(a)) the oscillations occur with a larger period.
The fact that it oscillates with a larger time period can be traced back to the behaviour 
of the fidelity at $\omega_{D}=11$. Indeed, for short times ($\omega_{D}=11.0$) the system stays in the initial ground 
state and after some time the fidelity starts to decrease, differently from the situation at $\omega_{D}=4.5$, where the 
system deviates from the initial state on much shorter time scales. 
Concerning the remaining tunneling modes, i.e. tunneling to higher energetical states that belong to the lowest-band (see discussion above),  
they are negligible as they provide a very small
contribution even for $\omega_{D}=\omega_D^c$. The latter has
already been seen in the last subsection, but it can also be
shown with the use of the multi-band analysis. 
On the other hand, Fig. 4(b) presents again the tunneling probability $|D_{\{N_i\};\textbf{I}}|^2$ for the energetically closest lowest-band states (i.e. same as Fig. 4(a)) when the driving
frequency is at resonance for different interaction strengths $g$.
For weak to intermediate interactions the tunneling amplitude
decreases and for strong interactions, e.g. $g=2.0$, a
destruction of the tunneling is observed for long time scales.

Now, let us consider the excitation dynamics. In this case it holds
$I_{m}^{(k)}\neq0$ for $k>1$. To this aim, we have analyzed the
probability of finding all the four bosons in the zeroth-band. The latter,  
can be expressed via eq.(7) as $|B_{\{N_i\};\textbf{I}}|^2=\sum_{\textbf{I}} |{}_{\textbf{I}}\left\langle
N_{1},N_{2},N_{3}| \Psi(t) \right\rangle|^2=\sum_{\textbf{I}}|C_{\{N_i\};\textbf{I}}|^2$, where the summation is performed over the excitation  
indices $\textbf{I}=(\textbf{$I_{L}$},\textbf{$I_{M}$}$,$\textbf{$I_{R}$})$ which, in terms of the multiband expansion, obey the constraints $I_{L}^{(1)}+I_{M}^{(1)}+I_{R}^{(1)}=N$ and
$I_{L}^{(j)}=I_{M}^{(j)}=I_{R}^{(j)}=0$ for all $j>1$. 
In particular, Fig. 5(a) shows the probability $|B_{\{N_i\};\textbf{I}}|^2$ for all the bosons to
reside in the zeroth-band for various driving frequencies
$\omega_{D}$ and a fixed amplitude $A=0.05$ during the
time evolution. At the
critical driving frequency a complete depopulation of the
zeroth-band at some specific time intervals is observed. In particular, this 
probability exhibits revivals, which are connected with the
enhancement of the (amplitude) oscillations of the single-particle
density [see also Fig. 2(b)]. On the other hand, for driving frequencies
different from the critical frequency the respective probability for all the bosons to
occupy the zeroth-band is rather large and is indeed 
dominant. However contributions from excited configurations cannot
be neglected, especially in the regions $7.0<\omega_{D}<8.0$ and
$10.0<\omega_{D}<15.0$, where the system significantly departs from
the initial state [see also Fig. 1(a) and Fig. 5(a) red dashed line]. Furthermore, Fig. 5(b)
presents the probability, at the critical driving frequency, to obtain a state
of $N_{0}\leq4$ particles in the first-excited band and the remaining to be
in the zeroth-band. The latter can be expressed as $|Q_{\{N_i\};\textbf{I}}|^2=\sum_{\textbf{I}}|C_{\{N_i\};\textbf{I}}|^2$ where the summation index $\textbf{I}=(\textbf{$I_{L}$},\textbf{$I_{M}$}$,$\textbf{$I_{R}$})$ obeys the constraints 
$I_{L}^{(1)}+I_{M}^{(1)}+I_{R}^{(1)}=N-N_{0}$,
$I_{L}^{(2)}=I_{M}^{(2)}=I_{R}^{(2)}=N_{0}$ and
$I_{L}^{(j)}+I_{M}^{(j)}+I_{R}^{(j)}=0$ for all $j>2$. Indeed,
the interplay between the four possible excitation scenarios from
the zeroth to the first excited-band (i.e. one-particle excitation, two-particle excitation etc) in the course of the dynamics is illustrated in
a transparent way. It is observed that the complete depopulation of
the zeroth-band is mainly accompanied by the excitation of three or
all the four bosons in the first-excited band. For long evolution
times the zeroth-band possesses a low population and states with one
or two bosons in the first excited-band are mainly populated. The
states with the most significant contribution are of the type
${\left| {1,2,1} \right\rangle _\textbf{I}}$ with
$I_{L}=I_{R}=(0,1,0)$ and $I_{M}=(0,2,0)$ or $I_{M}=(1,1,0)$.
We note that a small contribution comes from the state ${\left|
{1,2,1} \right\rangle _\textbf{I}}$ with $I_{L}=I_{R}=(0,1,0)$ and
$I_{M}=(0,1,1)$. This clearly shows that the most prominent
excitation process in our system originates from the energy
difference between each of the above states and ${\left| {1,2,1}
\right\rangle _I}$ with $I_{L}=I_{R}=(1,0,0)$ and $I_{M}=(2,0,0)$, 
namely the (initial) ground state configuration. 

Finally, in order to explore the impact of the
interactions on the dynamics, Fig. 5(c) shows the probability $|B_{\{N_i\};\textbf{I}}|^2$ for
long evolution times for all the bosons to be in the zeroth-band for
different interparticle repulsion at the driving frequency $\omega_{D}=\omega_{D}^c$. For the non-interacting case the
population of the zeroth-band shows revivals even for
long time scales, while, as the interaction strength is turned on, the
corresponding probability presents a decaying envelope. This envelope behaviour 
is a pure effect of the interactions and reflects also the initial
ground state configuration (see the discussion in Sec. II.C) which strongly depends on the interparticle interactions. As it can be seen for increasing
repulsion between the particles the probability for the system to remain in the zeroth-band, in the course of the dynamics, decays on increasingly shorter  
time scales and the system is dominated by different types of excitations, as expected intuitively.
\begin{figure}[ht]
        \centering
           \includegraphics[width=0.50\textwidth]{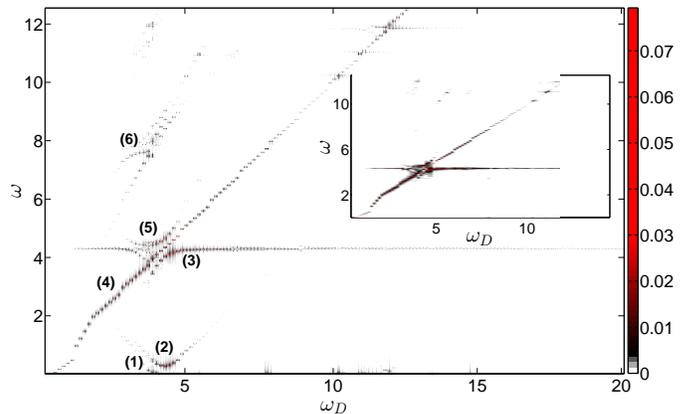}
                \caption{Local one-body density spectrum $\rho_{L}(\omega)$ (for the left well) 
                as a function of the driving frequency $\omega_{D}$ (measured in units of $\omega_{R}$). The driving amplitude has been chosen $A=0.05$. 
                Inset: the spectrum of the intra-well oscillations calculated via $\Delta\rho_{L}(t)$ (see also main text).}
\end{figure}
\begin{figure}[ht]
        \centering
           \includegraphics[width=0.50\textwidth]{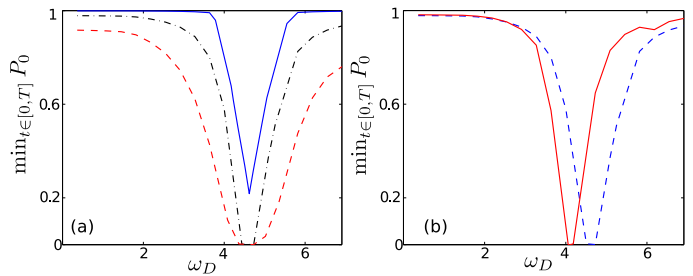}
                \caption{(a) Profile of the resonance for various driving amplitudes $A$= 0.01 (blue solid line), $A$= 0.05 (black dashed-dotted line) and $A$= 0.1 (red dashed line)
                obtained from the $\min_{t\in[0,T]}P_0(t)$ and $T$ being some fixed long evolution time, as a function of the driving frequency.
                (b) Same as (a) with $A=0.05$, but for different barrier heights $V_{0}$= 9.0 (red solid line) and $V_{0}$= 12.0 (blue dashed line).
                The system consists of four bosons confined in a triple-well with interparticle
                interaction $g=0.1$.}
\end{figure}

\begin{figure}[ht]
        \centering
           \includegraphics[width=0.50\textwidth]{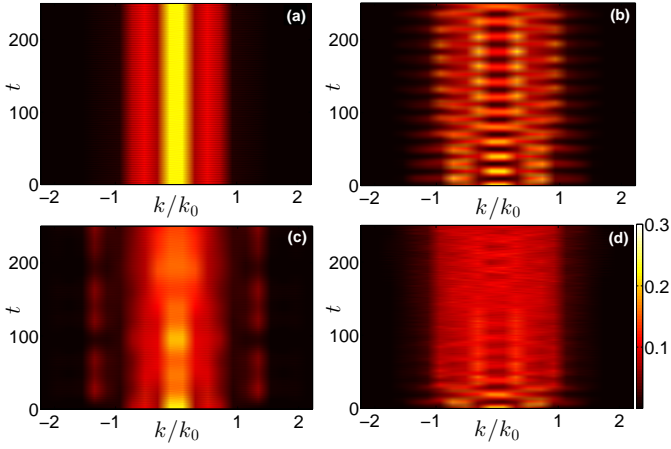}
                \caption{Momentum distribution of the one-body density as a function of time (measured in units of $\omega_{R}^{-1}$) for $g=0.1$ and
                different driving frequencies (a) before the critical frequency $\omega_{D}=2.0$, (b) at the critical
                frequency $\omega_{D}=\omega_D^c=4.5$ and (c) at $\omega_{D}=8.0$. (d) The case of strong interparticle repulsion for
                $g=2.0$ and $\omega_{D}=\omega_D^c$. The horizontal axis represents the lattice momenta in units of the inverse lattice vector $k_{0}=\pi/l$. In all panels $A=0.05$.}
\end{figure}

\subsection{Characteristics of the resonant behaviour}

To characterize the overall process with respect to the driving
frequency, we compute the spectrum of the local one-body density
\begin{equation}
\label{eq:8}\rho_{\alpha}(\omega) = \frac{1}{\pi}\int_{0}^{T}  dt\rho_{\alpha}(t) e^{i\omega t},
\end{equation}
where $\rho_{\alpha} (t) = \int_{d_{\alpha}}^{d'_{\alpha}} {d{x} \rho_{1}(x,t)}$ denotes 
the spatially over a single well integrated  
single-particle density at every time instant $t$. The index $\alpha=L,M,R$ corresponds to the left, middle or right well 
respectively, whereas the limits of the wells are denoted by $d_{\alpha}$, $d'_{\alpha}$. Note that in the present case all the  
components of $\rho_{\alpha}(\omega)$, i.e. $\rho_{L}(\omega)$, $\rho_{M}(\omega)$ and $\rho_{R}(\omega)$, are equivalent due to the considered large lattice depths and the 
employed driving scheme which enforces the bosons among different wells to oscillate in-phase. Figure 6 shows the above
spectrum, where five dominant branches (denoted as (1) to (5) in the Figure) can be observed. The lowest branch denoted as (1) in Fig.6 (in
the range $\omega\in[0,0.02]$) refers to the intraband tunneling
being restricted to the energetically closest number states e.g. from ${\left| {1,2,1} \right\rangle_\textbf{I}}$ ($I_{M}=(2,0,0)$, $I_{L}=I_{R}=(1,0,0)$) to 
${\left| {2,1,1} \right\rangle_\textbf{I}}$ ($I_{L}=(2,0,0)$, $I_{M}=I_{R}=(1,0,0)$). This branch is hardly visible in Fig.6 due to the 
presented wide range of frequencies that have been taken into account in order to visualize all 
the dynamical frequencies of the system. In addition, the next lowest
branch (denoted as (2)) at $\omega_{D}\in[4,5]$ and $\omega\in[0.05,1]$ corresponds
to the large amplitude density oscillations [see also Fig. 2(b)]. These mode frequencies 
have been already predicted via the Floquet analysis in Sec.
III.B [see Fig. 3(c) and (d)]. To investigate in some detail the
intra-well wavepacket dynamics the quantity
$\Delta\rho_{\alpha}(t)=\rho_{\alpha,1}(t)-\rho_{\alpha,2}(t)$ is
employed. Here, each well is divided from the center into two equal
parts, namely left and right, with $\rho_{\alpha,1}(t)$,
$\rho_{\alpha,2}(t)$ being the corresponding integrated densities at time
$t$. The index $\alpha=L,M,R$ stands for the left, middle and right
well, respectively. To determine the frequencies of this mode we
calculate the spectrum $\Delta {\rho _L}(\omega ) =
\frac{1}{\pi }\int {dt} \Delta {\rho _L}(t){e^{i\omega t}}$. The
inset of Fig. 6 presents the corresponding spectrum, thus showing the
emergent frequencies of the intra-well oscillations as a function of
the driving frequency $\omega_{D}$. We observe that the spectrum
$\Delta\rho_{L}(\omega)$ follows the evolution of the upper three
branches (denoted by (3), (4) and (5)) of the spectrum of $\rho_{L}(\omega)$, whereas in the region of
the resonance the intra-well oscillation measured via $\Delta\rho_{L}(t)$ features a beating dynamics, as expected. Hence, away from the region around the 
critical driving frequency the generated dipole mode possesses three
different frequencies, while close to $\omega_D^c$ the intra-well
dynamics come into a resonance. 
Therefore, one can induce this resonant intra-well dynamics by
adjusting the driving frequency. Finally, let us comment on the existence of some 
higher frequency components, e.g. branch (6) in Figure 6, which correspond to very fast 
intrawell oscillations (i.e. $\omega\ll\omega_{D}$) and possess a low amplitude (in comparison to the 
previous branches (1)-(5)).
\begin{figure*}[ht]
        \centering
           \includegraphics[width=0.80\textwidth]{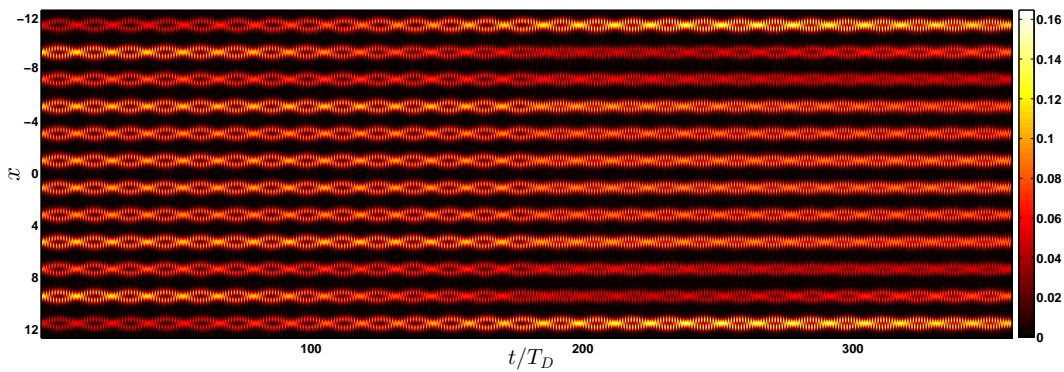}
                \caption{Time evolution of the one-body density $\rho_{1}(x,t)$ in a twelve-well
                potential for $\omega_{D}=4.5$. The
                driving amplitude is fixed to the value $A=0.05$, while the initial state corresponds to the ground state of five 
                weakly interacting bosons with $g=0.1$. The spatial extent of the lattice is expressed in units of $k_{0}^{-1}$, while the time units are rescaled in terms of the driving period $T_{D}$.}
\end{figure*}

In turn, we shall visualize the above mentioned resonance and
inspect how it depends on the lattice parameters. To this aim, the
minimal occupancy, during the evolution time $T$, of the zeroth-band $\min_{t\in [0,T]}
P_{0}(t)=\min_{t\in [0,T]}\sum_{\textbf{I}}|{}_{\textbf{I}}\left\langle N_{1},N_{2},N_{3}| \Psi(t)
\right\rangle|^2$, with the energetical indices
$I_{L}^{(1)}+I_{M}^{(1)}+I_{R}^{(1)}=N$ and
$I_{L}^{(j)}=I_{M}^{(j)}=I_{R}^{(j)}=0$ for every $j>1$ is used.
Employing the above quantity one can show that far from resonance
there are regions with non-negligible excitations i.e. $\min_{t\in [0,T]}P_{0}<1$
[e.g. at $\omega_{D}=11.0$ see also Fig. 1(a)] as well as regions
where $\min_{t\in [0,T]}P_{0}\approx1$ [e.g. $\omega_{D}=2.0$ in Fig. 1(a)]. Now
let us analyze the dependence of $\min_{t\in [0,T]}P_{0}$ on the driving
frequency around $\omega_{D}^c$. Firstly we study
the dependence of the resonance on the driving amplitude. 
In Figure 7(a) we show for an increasing driving amplitude the 
minimum of $\min_{t\in [0,T]}P_{0}$ as a function of the frequency $\omega_{D}$ which broadens and eventually reaches zero, meaning that the
zeroth-band has been complete depopulated [see also Fig. 5(a)]. On
the other hand, for small amplitudes the value of the minimum of $\min_{t\in [0,T]}P_{0}$
is non-zero and in the limit $A\to0$ its dependence on the
driving frequency disappears. Instead, in Fig. 7(b) we show how the minimal 
population of the zeroth-band ($\min_{t\in [0,T]}P_{0}$) varies as a function of the lattice
depth. For an increasing lattice depth it is known that the energy
gaps among the different energy levels become larger. This
phenomenon can intuitively be understood in terms of a tight-binding
approximation. For simplicity let us assume only a nearest neighbour
coupling $J\propto\int dx
W_{s}(x)[\frac{p^{2}}{2m}+V_{0}\sin^{2}(x)]W_{s+1}(x)$ between
the sites $s$ and $s+1$, where $W_{s}(x)$ are the on-site localized
Wannier states. Then, within this approximation, which is valid for
a relatively deep potential, the resulting eigenvalues are
$E_{k-1}=E_{0}^{on-site}-2J\cos(\frac{k\pi} {N+1})$ ($k=1,2,...,N$),
where $E_{0}^{on-site}$ are the on-site energies. Thus, the
resonance can be tuned at will, i.e. for a decreasing lattice depth
the $\omega_D^c$ is negatively shifted, as it is confirmed by the
numerical results of Fig. 7(b). Finally, let us comment on the dependence of the position of the resonance 
on the interparticle interaction strength $g$. Indeed, in order to investigate whether there is such a dependence various interaction strengths    
(for the same particle number $N=4$), e.g. $g=0.1$, $g=1.0$ and $g=3.0$, have been considered (omitted here for brevity) and it was found that the position 
of the resonance is essentially unaffected.

In the following, let us inspect the momentum distribution with varying 
driving frequency with the aim of understanding whether signatures of a parametric
amplification of matter-waves can be observed. The momentum
distribution is a routinely employed observable in atomic quantum
gases experiments as it is accessible via time-of-flight
measurements \cite{Bloch}. This quantity can be calculated as the
Fourier transformation of the one-body reduced density matrix as
\begin{equation}
\label{eq:9}n(k,t)= \frac{1}{2\pi }\int \int
{dx}{dx'}\rho_{1}(x,x';t)e^{-ik(x-x')}.
\end{equation}
Here $\rho_{1}(x,x';t)$ denotes the one-body reduced density matrix, being obtained by tracing out all the bosons but one
in the density of the $N$-body system. The panels 8(a)-(c) of Fig. 8 present the time evolution of the momentum distribution 
for different driving frequencies before, on, and after the resonance. As it can be noted, exactly
at the resonance the momentum distribution exhibits a special
pattern, that is, some additional lattice momenta are periodically
activated during the dynamics. In particular,
it is observed that the modes $\pm\frac{k_{0}}{2}\simeq\pm1.57$, $\pm{k_{0}}\simeq\pm3.14$, $\pm\frac{3k_{0}}{2}\simeq\pm4.713$ are populated, whereas out
of resonance only the $\pm{k_{0}}$ modes are significantly populated. 
The population of the $\pm{k_{0}/2}$, $\pm3{k_{0}/2}$ modes at $\omega_D = \omega_D^c$ is reminiscent of the parametric amplification of 
matter-wave phenomenon, as observed experimentally in ref. \cite{Gemelke}. However, an exact correspondence with ref. \cite{Gemelke} cannot be made 
due to the very different setup of our system, i.e. its finite size and the hard wall boundaries. A detailed study of this 
process, also for higher particle numbers and lattice potentials, would be desirable, but it is clearly beyond the scope of this work.
Furthermore, Fig. 8(d) shows the momentum distribution at resonance, but for a strong interparticle repulsion $g=2.0$. 
The expected periodic pattern for large evolution times is blurred as an effect of the strong interaction which decreases the degree of coherence.
 
Finally, in order to demonstrate that our findings are of general character we investigate a larger
lattice system with a filling factor smaller than unity. Specifically, 
the case of five bosons in a twelve-well finite lattice has been 
considered. Concerning the ground state with filling factor $\nu<1$,
the most important aspect is the spatial redistribution of the atoms
as the interaction strength increases. Indeed, as the repulsion
increases from the non-interacting to the weak interaction regime
the atoms are pushed from the central to the outer sites which gain and lose
population in the course of increasing $g$.

In the following, the shaking dynamics applied at $t=0$ to the ground
state of the five bosons which are trapped in the twelve-well potential in the weak interaction regime ($g=0.1$) is explored. The
emergent non-equilibrium behaviour shows similar characteristics as
in the previous setup with filling $\nu>1$, i.e. the occurrence of an intrawell dipole and an interwell 
tunneling mode. Interestingly, at
the same frequency $\omega_{D}=\omega_D^c=4.5$ a resonance of the
intra-well dynamics is observed. Figure 9 presents the one-body density evolution
exactly at the critical point $\omega_D$. As in the case for setups with
filling $\nu>1$, the formation of enhanced density oscillations at
each site is observed, which is in relation to the time periods where 
the zeroth-band is completely depopulated during the evolution.
Employing a corresponding number state analysis the significant contribution of two kinds of
number states has been confirmed: a) Either
$I_{1}^{(1)}+...+I_{12}^{(1)}=N-1$, $I_{1}^{(3)}=...=I_{12}^{(3)}=0$
and one with $I_{k}^{(2)}=1$ for $k=1,...,12$ or b)
$I_{1}^{(1)}+...+I_{12}^{(1)}=N-1$, $I_{1}^{(2)}=...=I_{12}^{(2)}=0$
and a certain $I_{k}^{(3)}=1$ for $k=1,...,12$. Notice that the same kind of number states have been found to contribute significantly also 
in the dynamics of four bosons in the triple-well. The above mentioned observations suggest a generalization of 
the observed phenomena to larger systems as well. Indeed, the same shaken scheme has been tested in different systems (omitted here for brevity), 
e.g. 10 bosons in a triple-well, 6 bosons in five wells etc, confirming that the above observed resonant-like behaviour of the bosonic ensemble occurs in each setup.

\section{Conclusions and Outlook}

The correlated non-equilibrium quantum dynamics of few-body bosonic ensembles
induced by the driving of a finite-size optical lattice has been
investigated. Our work focuses particularly on the regimes of large lattice depths
and small driving amplitudes. This choice has been made in
order to limit the degree of excitations that would otherwise lead to
heating processes. Starting from the ground state of a weak or
strongly interacting small ensemble, we have examined in detail the
time evolution of the system induced by periodically driving the
optical lattice. We find that the dynamical evolution of the system
is governed by two main modes: the inter-well tunneling and the
intra-well dipole-like mode. The dynamical behaviour of the system in
the non-interacting regime has been firstly analyzed via Floquet
theory, that is, at the single-particle level, providing an accurate
interpretation of the observed processes. For large particle numbers
and large interaction strengths, however, such a single-particle
description was not anymore sufficient to provide an exhaustive
explanation of the observed dynamics, and a multi-band Wannier
number state expansion has been employed.

The inter-well tunneling mode is weak as a consequence of the deep optical lattice
and the small driving amplitude. On the other hand, the local dipole mode has been
identified from the intra-well oscillations of bosons in the
individual wells. Remarkably enough, it has been found that by tuning the driving frequency the intra-well dynamics
experiences a resonant-like behaviour. This is manifested e.g. by the enhanced oscillations in the
one-body density evolution or from the periodic population of additional lattice momenta in the momentum distribution
of the one-body density. Additionally, on a single-particle level in terms of Floquet theory, it has been shown that 
in the proximity of the resonance the first two FMs possess the main contribution, while away from resonance the dynamics 
can be described with the inclusion of the first FM. To explain the enhanced population of the second FM at resonance the corresponding quasienergy 
spectrum has been employed, revealing avoided-crossings between the first two FMs at certain driving frequencies. To obtain the frequencies 
which refer to the on-site and tunneling dynamics, the corresponding frequencies associated with the interference terms between the FMs have been employed showing 
pronounced on-site oscillations and an enhancement of the inter-well tunneling mode in the vicinity of the resonance. Considering an ensemble of few-bosons we examined 
the influence of the interatomic interactions both for the inter- and intra-well generated modes. Indeed, it has been found that the repulsion 
affects each of the aforementioned modes, yielding a destruction of the inter-well tunneling for strong interactions and an enhancement of the excitations (i.e. the contribution of higher-band states).
Moreover, in the spectrum of the local one-body density with respect to the driving frequency all the relative
dynamical frequencies, e.g. on-site oscillations and tunneling period have been identified. Finally, the occurrence of the above resonance seems to be universal in a periodically driven lattice as it is independent
of the filling factor, the boundary conditions or the interparticle repulsion.

We would like to underline that, contrarily to related studies based e.g. on
effective model Hamiltonians or lattice calculations with tensor network methods, our many-body analysis based
on the ab initio MCTDHB method has the advantage to provide the complete system wavefunction in space and time. Thus, it 
enables us to accurately identify the involved intra- and inter-well band excitations.

Let us comment on possible future investigations. Although
in the present work we did not employ the multi-layer structure of
the ML-MCTDHB method, our ab-initio approach is well suited to describe
the dynamics of multi bosonic species. Given this, a first natural
extension would be to study the driven dynamics of mixtures
consisting of different bosonic species in order to unravel the
induced excitation modes or to device schemes for selective
transport of an individual bosonic component. In relation to the present
study, it would be interesting to simulate the parametrical
amplification of matter-waves with interesting
applications, like the generation of four-wave mixing, entanglement
production, but also for fundamental tests of quantum mechanics with
massive particles like the Hong-Ou-Mandel experiment, as recently
performed with a Bose-Einstein condensate \cite{Lopes}.

\appendix

\section{Appendix A: The Computational Method: ML-MCTDHB and MCTDHB}

Our computational approach to solve the many-body Schr\"{o}dinger equation of the interacting bosons relies on the Multi-Layer MultiConfiguration
Time-Dependent Hartree method for Bosons (ML-MCTDHB)
\cite{Cao,Kronke} which constitutes an ab-initio method for the calculation of
stationary properties and in particular the non-equilibrium quantum
dynamics of bosonic systems of different species. For a single species it reduces to
MCTDHB which has been established in refs. \cite{Alon,Alon1,Streltsov} and
applied extensively \cite{Streltsov,Streltsov1,Alon2,Alon3}. The wavefunction is represented by a set of
variationally optimized time-dependent orbitals which implies
an optimal truncation of the Hilbert space by employing a
time-dependent moving basis where the system can be
instantaneously optimally represented by the corresponding time-dependent permanents.
To be self contained let us briefly introduce the basic concepts of the method
and discuss the main ingredients of our implementation.

Within the MCTDHB method the time-dependent Schr\"{o}dinger equation
$\left( {i\hbar {\partial _t} - H} \right)\Psi (x,t) = 0$ is solved
as an initial value problem $\ket{{\Psi (0)}} = \left| {{\Psi _0}}
\right\rangle$. The many-body wavefunction which is expanded in terms of
the bosonic number states $\left| {{n_1},{n_2},...,{n_M};t}
\right\rangle$, based on time-dependent single-particle
functions (SPFs) $\left| \phi_{i}(t) \right\rangle$, $i=1,2,...,M$,
reads
\begin{equation}
\label{eq:10}\left| {\Psi (t)} \right\rangle  = \sum\limits_{\vec n
} {{C_{\vec n }}(t)\left| {{n_1},{n_2},...,{n_M};t} \right\rangle }.
\end{equation}
Here $M$ is the number of SPFs and the summation $\vec n$ is over
all the possible particle combinations $n_{i}$ such that the total number
of bosons is conserved and equal to $N$. To determine the time-dependent wave function $\left|
\Psi(t) \right\rangle$ we need the equations of motion for the
coefficients ${{C_{\vec n }}(t)}$ and of the SPFs $\left| \phi_{i}(t) \right\rangle$.
Following the Dirac-Frenkel
\cite{Frenkel,Dirac} variational principle i.e. ${\bra{\delta
\Psi}}{i{\partial _t} - \hat{ H}\ket{\Psi }}=0$ we end up with the
well-known MCTDHB equations of motion
\cite{Alon,Streltsov,Alon1,Broeckhove} consisting of a set of $M$
non-linear integrodifferential equations of motion for the orbitals
which are coupled to the $\frac{(N+M-1)!}{N!(M-1)!}$ linear
equations of motion for the coefficients.

For our numerical implementation a discrete variable representation (DVR) for the
SPFs and a sin-DVR, which intrinsically introduces hard-wall
boundaries at both edges of the potential, has been employed. The
preparation of the initial state has been performed by using the so-called
relaxation method in terms of which one obtains the lowest
eigenstates of the corresponding $m$-well setup. The key idea is to
propagate some trial wave function ${\Psi ^{(0)}}(x)$ by the
non-unitary operator ${e^{ - H\tau }}$. This is equivalent to an imaginary time
propagation and for $\tau  \to \infty $, the propagation converges to
the ground state, as all other contributions (i.e., $e^{-E_n\tau}$) are 
exponentially suppressed. In turn, we periodically drive the optical lattice and study the
evolution of $\Psi ({x_1},{x_2},..,{x_N};t)$ in the $m$-well
potential within MCTDHB. To ensure the convergence of our
simulations we have used up to 9 single
particle functions thereby observing a systematic convergence of our
results for sufficiently large spatial grids. An additional
criterion that confirms the achieved convergence is the population of the
lowest occupied natural orbital kept in each case below $0.1\%$.

\section*{Acknowledgments}
S.M. thanks the Hamburgisches Gesetz zur F{\"o}rderung des wissenschaftlichen
und k{\"u}nstlerischen Nachwuchses (HmbNFG) for a PhD Scholarship.
P.S gratefully acknowledges funding by the Deutsche Forschungsgemeinschaft (DFG) in the framework of the
SFB 925 ''Light induced dynamics and control of correlated quantum
systems''. A.N. gratefully acknowledges discussions with
Klaus M{\o}lmer related to four-wave mixing.

{}


\begin{thebibliography}{60}

\bibitem{Goldman}N. Goldman, and J. Dalibard, Phys. Rev. X \textbf{4}, 031027 (2014).

\bibitem{Goldman1} N. Goldman, J. Dalibard, M. Aidelsburger, and N. R. Cooper,
Phys. Rev. A \textbf{91}, 033632 (2015).

\bibitem{Morsch1}O. Morsch, and M. Oberthaler, Rev. Mod. Phys. \textbf{78}, 179 (2006).

\bibitem{Bloch}I. Bloch, J. Dalibard, and W. Zwerger,  Rev. Mod. Phys. \textbf{80}, 885 (2008).

\bibitem{Olshanii}M. Olshanii, Phys. Rev. Lett. \textbf{81},
938 (1998).

\bibitem{Grimm}R. Grimm, M. Weidem{\"u}ller, and Y. B. Ovchinnikov, Adv. At. Mol. Opt. Phys. \textbf{42},
95-170 (2000).

\bibitem{Santos}L. Santos, M. A. Baranov, J. I. Cirac, H. U. Everts, H. Fehrmann, and M. Lewenstein,
Phys. Rev. Lett. \textbf{93}, 030601 (2004).

\bibitem{Inouye}S. Inouye, J. Goldwin, M. L. Olsen, C. Ticknor, J. L. Bohn,
and D. S. Jin, Phys. Rev. Lett. \textbf{93}, 183201 (2004).

\bibitem{Kohler} T. K{\"o}hler, K. Goral, and P. S. Julienne, Rev. Mod. Phys. \textbf{78},
1311 (2006).

\bibitem{Chin}C. Chin, R. Grimm, P. Julienne, and E. Tiesinga, Rev. Mod. Phys. \textbf{82},
1225 (2010).

\bibitem{Lewenstein1} M. Lewenstein, A. Sanpera, and V. Ahufinger, Ultracold Atoms in Optical Lattices:
Simulating quantum many-body systems, (Oxford University Press, 2012).

\bibitem{Choi}D. I. Choi, and Q. Niu, Phys. Rev. Lett. \textbf{82}, 2022 (1999).

\bibitem{Dahan} M. B. Dahan, E. Peik, J. Reichel, Y. Castin, and C. Salomon,
Phys. Rev. Lett. \textbf{76}, 4508 (1996).

\bibitem{Morsch}O. Morsch, J. H. M{\"u}ller, M. Cristiani, D. Ciampini, and E. Arimondo,
Phys. Rev. Lett. \textbf{87}, 140402 (2001).

\bibitem{Peik}E. Peik, M. B. Dahan, I. Bouchoule, Y. Castin, and C. Salomon, Phys. Rev. A \textbf{55}, 2989 (1997).

\bibitem{Cristiani} M. Cristiani, O. Morsch, J. H. M{\"u}ller, D. Ciampini, and E. Arimondo, Phys. Rev. A \textbf{65}, 063612 (2002).


\bibitem{Wilkinson} S. R. Wilkinson, C. F. Bharucha, K. W. Madison, Q.
Niu, and M. G. Raizen, Phys. Rev. Lett. \textbf{76}, 4512 (1996).

\bibitem{Niu} Q. Niu, X. G. Zhao, G. A. Georgakis, and M. G. Raizen,
Phys. Rev. Lett. \textbf{76}, 4504 (1996).


\bibitem{Sias} C. Sias, H. Lignier, Y. P. Singh, A. Zenesini, D. Ciampini,
O. Morsch, and E. Arimondo, Phys. Rev. Lett. \textbf{100}, 040404
(2008).

\bibitem{Eckardt} A. Eckardt, C. Weiss, and M. Holthaus,   Phys. Rev. Lett. \textbf{95}, 260404 (2005).

\bibitem{Gemelke} N. Gemelke, E. Sarajlic, Y. Bidel, S. Hong, and S. Chu, Phys. Rev. Lett. \textbf{95}, 170404 (2005).

\bibitem{Hilligsoe} K. M. Hilligs{\o}e, and K. M{\o}lmer, Phys. Rev. A \textbf{71}, 041602 (2005).

\bibitem{Lopes} R. Lopes, A. Imanaliev, A. Aspect, M. Cheneau, D. Boiron, and C. I. Westbrook, Nature \textbf{520} (2015).

\bibitem{Zheng}W. Zheng, and H. Zhai, Phys. Rev. A \textbf{89}, 061603 (2014).

\bibitem{Lignier} H. Lignier, C. Sias, D. Ciampini, Y. Singh, A. Zenesini, O. Morsch, and E. Arimondo,  Phys. Rev. Lett. \textbf{99}, 220403 (2007).

\bibitem{Struck} J. Struck, C. {\"O}lschl{\"a}ger, M. Weinberg, P. Hauke, J. Simonet,
A. Eckardt, M. Lewenstein, K. Sengstock and P. Windpassinger, Phys.
Rev. Lett. \textbf{108}, 225304 (2012).

\bibitem{Parker} C. V. Parker, L. C. Ha, and C. Chin, Nature Phys. \textbf{9}, 769-774 (2013).

\bibitem{Choudhury} S. Choudhury, and E. J. Mueller, Phys. Rev. A \textbf{90}, 013621 (2014).

\bibitem{Strater} C. Str{\"a}ter, and A. Eckardt, Phys. Rev. A \textbf{91}, 053602 (2015).

\bibitem{Iucci} A. Iucci, M. A. Cazalilla, A. F. Ho, and T. Giamarchi, Phys. Rev.
A \textbf{73}, 041608 (2006).

\bibitem{Schneider} P. I. Schneider, and A. Saenz, Phys. Rev. A \textbf{85}, 050304 (2012).

\bibitem{Alon} O. E. Alon, A. I. Streltsov, and L. S. Cederbaum,  J. Chem. Phys. \textbf{127}, 154103 (2007).

\bibitem{Alon1}O. E. Alon, A. I. Streltsov, and L. S. Cederbaum, Phys. Rev. A \textbf{77}, 033613 (2008).

\bibitem{Kim}J. I. Kim, V. S. Melezhik, and P. Schmelcher, Phys. Rev. Lett. \textbf{97}, 193203 (2006).

\bibitem{Giannakeas}P. Giannakeas, F. K. Diakonos, and P. Schmelcher,  Phys. Rev. A \textbf{86}, 042703 (2012).

\bibitem{Mistakidis} S. I. Mistakidis, L. Cao, and P. Schmelcher, J. Phys. B: At. Mol. Opt. Phys. \textbf{47}, 225303 (2014).

\bibitem{Mistakidis1} S. I. Mistakidis, L. Cao, and P. Schmelcher, Phys. Rev. A \textbf{91}, 033611 (2014).

 \bibitem{Gorin}T. Gorin, T. Prosen, T. H. Seligman, and M. \v{Z}nidari\v{c}, Phys. Rep. \textbf{435}, 33 (2006).

 \bibitem{Tannor} D.J. Tannor, \textit{Introduction to Quantum Mechanics: A Time-Dependent Perspective}, (University Science Books, Sausalito, California, 2007).
\bibitem{Hanggi} M. Grifoni and P. H\"{a}nggi, Phys. Rep. {\bf 304}, 229-354 (1998).
\bibitem{Wulf:2014} T. Wulf, C. Petri, B. Liebchen and P. Schmelcher, Phys. Rev. E {\bf 90}, 042913 (2014).
\bibitem{Wulf:2015} T. Wulf, B. Liebchen and P. Schmelcher, Phys. Rev. A {\bf 91}, 043628 (2015).

\bibitem{Cao} L. Cao, S. Kr{\"o}nke, O. Vendrell, and P. Schmelcher,  J. Chem. Phys. \textbf{139}, 134103 (2013).

\bibitem{Kronke}S. Kr{\"o}nke, L. Cao, O. Vendrell, and P. Schmelcher, New J. Phys. \textbf{15}, 063018 (2013).

\bibitem{Streltsov} A. I. Streltsov, O. E. Alon, and L. S. Cederbaum, Phys. Rev. Lett. \textbf{99}, 030402 (2007).

\bibitem{Streltsov1} A. I. Streltsov, K. Sakmann, O. E. Alon, and L. S. Cederbaum, Phys. Rev. A \textbf{83}, 043604 (2011).

\bibitem{Alon2} O. E. Alon, A. I. Streltsov, and L. S. Cederbaum, Phys. Rev. A \textbf{76}, 013611 (2007).

\bibitem{Alon3} O. E. Alon, A. I. Streltsov, and L. S. Cederbaum, Phys. Rev. A \textbf{79}, 022503 (2009).

\bibitem{Frenkel} J. Frenkel, in Wave Mechanics 1st ed. (Clarendon Press, Oxford, 1934), pp. 423-428.

\bibitem{Dirac} P. A. Dirac, (1930, July). Proc. Camb. Phil. Soc.
(Vol. \textbf{26}, No. 03, pp. 376-385). Cambridge University Press.

\bibitem{Broeckhove} J. Broeckhove, L. Lathouwers, E. Kesteloot, and P. Van Leuven, 
Chem. Phys. Lett. \textbf{149}, 547 (1988).




\end{thebibliography}
\end{document}